\documentclass{article}

\usepackage{PRIMEarxiv}

\usepackage[utf8]{inputenc} 
\usepackage{amsmath}
\usepackage[T1]{fontenc}    
\usepackage{hyperref}       
\usepackage{url}            
\usepackage{booktabs}       
\usepackage{amsfonts}       
\usepackage{nicefrac}       
\usepackage{microtype}      
\usepackage{lipsum}
\usepackage{fancyhdr}       
\usepackage{graphicx}       
\graphicspath{{media/}}     
\usepackage[round]{natbib}
\usepackage{url}
\hypersetup{
  colorlinks   = true, 
  urlcolor     = blue, 
  linkcolor    = blue, 
  citecolor   = blue 
}
\usepackage{authblk}

\usepackage{caption}
\usepackage{subcaption}

\newcommand{\fof}{\texttt{FoF}}
\newcommand{\latent}{\texttt{Latent}}

\usepackage{algorithm2e}
\RestyleAlgo{ruled}
\usepackage{lipsum}
\usepackage{algorithmic}
\makeatletter
\renewcommand{\Indentp}[1]{%
  \advance\leftskip by #1
  \advance\skiptext by -#1
  \advance\skiprule by #1}%
\renewcommand{\Indp}{\algocf@adjustskipindent\Indentp{\algoskipindent}}
\renewcommand{\Indm}{\algocf@adjustskipindent\Indentp{-\algoskipindent}}
\makeatother

\title{Delayed and Indirect Impacts of Link Recommendations
}

\newcommand\blfootnote[1]{%
  \begingroup
  \renewcommand\thefootnote{}\footnote{#1}%
  \addtocounter{footnote}{-1}%
  \endgroup
}

\author[1]{Han Zhang}
\author[2]{Shangen Lu}
\author[3]{Yixin Wang}
\author[1]{Mihaela Curmei}
\affil[1]{University of California, Berkeley}
\affil[2]{The Wharton School}
\affil[3]{University of Michigan}

\begin{document}
\maketitle
\blfootnote{$^1$Corresponding author: mcurmei@berkeley.edu}

\begin{abstract}
The impacts of link recommendations on social networks are challenging to evaluate, and so far they have been studied in limited settings. Observational studies are restricted in the kinds of causal questions they can answer and  naive A/B tests often lead to biased evaluations due to unaccounted network interference. Furthermore, evaluations in simulation settings are often limited to static network models that do not take into account the potential feedback loops between link recommendation and organic network evolution. To this end, we study the impacts of recommendations on social networks in dynamic settings. Adopting a simulation-based approach, we consider an explicit dynamic formation model---an extension of the celebrated Jackson-Rogers model---and investigate how link recommendations affect network evolution over time. Empirically, we find that link recommendations have surprising \emph{delayed} and \emph{indirect} effects on the structural properties of networks. Specifically, we find that link recommendations can exhibit considerably different impacts in the immediate term and in the long term. For instance, we observe that friend-of-friend recommendations can have an immediate effect in decreasing degree inequality,  but in the long term, they can make the degree distribution substantially more unequal.  
Moreover, we show that the effects of recommendations can persist in networks, in part due to their indirect impacts on natural dynamics even after recommendations are turned off. We show that, in counterfactual simulations, removing the indirect effects of link recommendations can make the network trend faster toward what it would have been under natural growth dynamics.
\end{abstract}


\keywords{recommendation systems \and link recommendation \and dynamic graph modeling \and social dynamics in graphs}

\section{Introduction}

Link recommendation algorithms such as Facebook's "People You May Know", Twitter's "Who to Follow" and LinkedIn's "Recommended for You" have an ever-increasing influence on the evolution of social networks, with some accounts crediting over 50\% of links to algorithmic recommendations \citep{PeopleYouMayKnow}.
This can cause downstream effects on information flow, opinion dynamics, and resource allocation. For instance, recommendations can be a polarizing force increasing network segregation, which in turn may re-inforce opinion echo-chambers \citep{cinus2022effect} or restrict access to information and resources for less connected communities \citep{dandekar_biased_2013, auletta_impact_2022, santos_link_2021}. At the same time, recommendation systems can also surface “long-range” connections between nodes that would not otherwise be exposed to each other, and thus, increase network integration by promoting and maintaining diverse links \citep{rajkumar_causal_2022, dangelo_recommending_2019}. Given the ubiquity of algorithmic recommendations on social media, studying their impacts is seeing increased academic and regulatory interest. However, such studies are challenging for a variety of reasons, such as lack of normative framing \citep{daly2010network} and limited access to large-scale platforms. In this work, we explore a more foundational evaluation challenges stemming from the fact that social networks have underlying dynamics that interfere with algorithmic recommendations.

Existing real-world evaluations of link recommendation algorithms rely on A/B tests (experimental) or longitudinal data (observational). However, both experimental and observational evaluations can yield misleading conclusions. The validity of A/B tests relies on the Stable Unit Treatment Value Assumption (SUTVA) \citep{imbens_causal_2015, gui2015network}, which is violated in  case of network interference. Observational studies may fail to assess causal impacts as longitudinal evaluation lacks counterfactual measurements on what the network evolution would have been without algorithmic recommendations. These challenges have motivated the use of simulation-based evaluations of link recommendations. In simulation studies, one can make explicit network modeling assumptions and then evaluate the impact of recommenders. Despite a growing number of works in this space \citep{stoica_algorithmic_2018, cinus2022effect, fabbri_effect_2020, fabbri2022exposure, ferrara2022link, santos_link_2021}, 
existing simulation-based evaluation falls short of providing insights into the mechanism through which recommendations impact social networks. Existing simulation studies primarily consider static networks, and thus do not take into account feedback loops between link recommendation and organic network evolution.



In dynamic settings, the main challenge is to measure impacts relative to a "baseline" network. Given this, existing works often measure the effects relative to the initial networks; they are rarely measured relative to a plausible counterfactual based on the natural evolution of the graph without link recommendations. Such evaluations can lead to qualitatively wrong conclusions. For instance, \citet{abebe_effect_2022} shows that triadic closure -- the most common type of friend-of-friend recommendations -- can reduce segregation with respect to the initial network before intervention. However, triadic closure can, at the same time, increase network segregation in relative terms with respect to a natural evolution dynamic which assumes the addition of random edges.

\begin{figure}
    \centering
    \includegraphics[trim={11cm 2cm 6cm 5cm}, clip, width=0.85\linewidth]{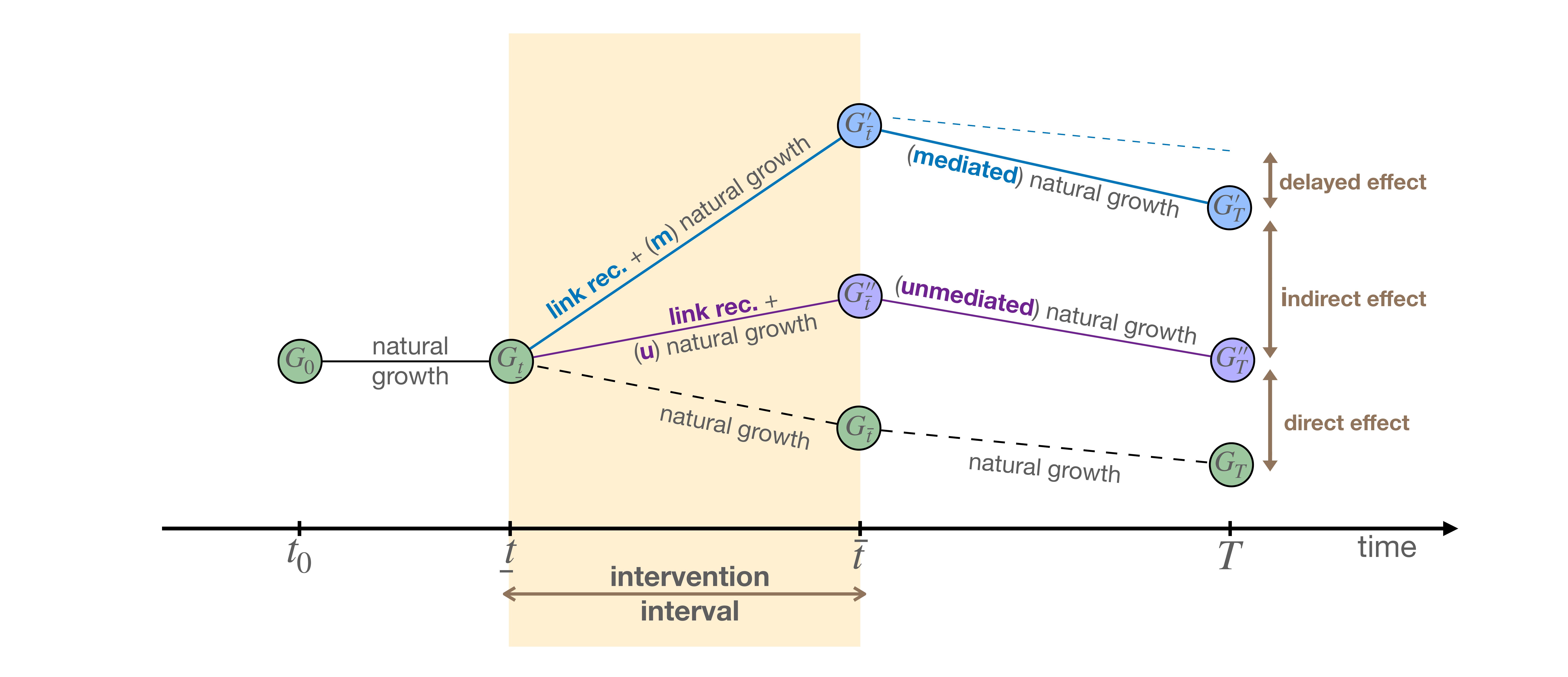}
    \caption{Delayed and indirect effects: The image displays two counterfactual evolutions of the network. The solid blue line represents the trajectory for an intervention interval $[\underline{t}, \overline{t}]$ in which the full network receives algorithmic recommendations. The dashed black trajectory corresponds to a counterfactual network evolution without recommendations. The total causal effect of recommendation at time $t$; $\text{Effect}_t$ can be computed as the difference between the counterfactual trajectories at time $t$ (solid blue line and dashed black line). The delayed effect at some time $T\geq \overline{t}$ is the difference:  $\text{Effect}_T - \text{Effect}_{\overline{t}}$. 
 The solid purple trajectory illustrates the counterfactual evolution of the network in which the indirect effects are removed. The difference between the purple and dashed curve captures the direct effects and the difference between the blue and purple line captures the indirect effects.} 
    \label{fig:sketch_effects}
\end{figure}

In this work, we initiate a study to explore \emph{dynamic} impacts of link recommendations through simulations. We find that link recommendations can have surprising delayed and indirect effects on the structural properties of networks. 
For instance, the effects of friend-of-friend recommendations can vary in the short-term and long-term: it may alleviate degree inequality   in the short term, but increase degree inequality in the long term;  the perceived increase in degree inequality can even be significantly more pronounced after the recommendations are discontinued. Moreover, we demonstrate that the effects of link recommendations can persist in the network even after recommendations have been discontinued. This phenomenon is due to the fact that recommendations impact the network in two ways: directly through the creation of algorithmic edges, and indirectly by influencing the natural growth dynamics. Indirect effects amplify the direct effects, contributing to the persistent impact of link recommendation algorithms. A stylized illustration of the indirect and delayed effects can be found in Figure~\ref{fig:sketch_effects}.


\paragraph{Contributions.}
We enumerate our contribution along modeling, evaluation, and experimental findings.
\begin{itemize}
\item\emph{Modeling:} We propose a dynamic network formation model which extends upon the Jackson-Rogers model~\citep{jackson2007meeting} to incorporate algorithmic recommendations. Unlike the classic model, our dynamic model includes not only the  original two phases ---dubbed "meeting strangers" and "meeting friends"---but also a third phase of ``meeting recommendations.'' Additionally, we consider latent node representations, enabling us to model community structure and node activity levels in a flexible manner.
\item \emph{Evaluation:} 
We monitor the progression of network metrics over different intervention windows. We compare the immediate impacts observed during intervention with the delayed impacts observed after the intervention has ended. Further, we measure the indirect impacts that recommendations have on network properties. To tease out the indirect impact, we compare the observed network evolution with a counterfactual baseline network that discounts the influence of recommendations on natural growth.
\item \emph{Experimental findings:} 
Our study reveals diverse qualitative patterns for delayed and indirect effects; we find that different durations of the intervention and/or different times of measurement can lead to drastically different conclusions about \emph{"How do recommendations impact networks?."} Furthermore, we find that indirect effects can be substantial; they can significantly amplify the impact of recommendations and persist even after the intervention has ended.
\end{itemize}

\section{Related Work}\label{sec:related}
\paragraph{Simulation studies.}
Simulation studies typically analyze the impact of link recommendations on static networks. Some works focus on a single round and examine biases observed in proposed recommendations, finding that homophily --- the preference for within-group links --- leads to exposure bias toward the more homophilous group, even if it's a minority \citep{karimi2018homophily, fabbri_effect_2020, espin2022inequality, stoica_algorithmic_2018}.  Similarly, \citep{fabbri2022exposure, ferrara2022link} find that homophily and node degree are strong indicators of which nodes will receive disproportionate visibility. Other works make explicit behavioral assumptions on how nodes accept link recommendations and consider cumulative effects of link recommendations over multiple rounds. These studies reveal algorithmic amplification of biases via "rich get richer" effects and increases in observed homophily over time \citep{fabbri2022exposure, ferrara2022link}. A shortcoming of existing simulation based evaluations is that they implicitly assume that the addition of algorithmic edges is the only change in the network. Conversely, in our work we make explicit modelling assumptions about the underlying network dynamics. This evaluation setup allows us to measure the impact of link recommendations with respect to counterfactual natural evolution. Furthermore, by emphasizing underlying temporal dynamics we can pose more subtle evaluation questions such as: "How does the effect of algorithmic intervention fade over time once recommendations are stopped?" or "How do algorithmic recommendations bias the underlying network growth?". 
\paragraph{Platform studies.}
There is limited publicly available research evaluating link recommendation algorithms on real social networking platforms. In one experimental study \citep{daly2010network}, several recommendation algorithms were compared on IBM's SocialBlue network and found to reduce group homophily. The study also revealed that friend-of-friend recommendations had the highest rate of acceptance but the lowest level of edge activity. On the other hand, \citep{rajkumar_causal_2022} found that recommending more distant connections or "weak ties" through LinkedIn's "People You May Know" algorithm led to higher transmission of job opportunities.
In observational settings, a longitudinal study comparing the Twitter network before and after the introduction of the "Who To Follow" recommender in 2010 \citep{su2016effect} found that while recommendations increased the number of connections for all users, the highest gains were achieved by the most popular nodes. Conversely, a study comparing links formed naturally and links formed via recommendations on Flickr and Tumblr \citep{aiello_evolution_2017} found that the recommended links were more diverse and less biased towards popular users. So far, existing observational platform studies provide limited understanding of the underlying mechanisms and lack access to counterfactual network evolution. A/B tests, on the other hand, may produce wrong estimates of the effects due to network interference. Our counterfactual simulations allow us to articulate in stylized settings the source of bias in both longitudinal and A/B evaluations.
\paragraph{Theoretical investigations.}
The interaction between homophily and friend-of-friend recommendations was studied theoretically in a number of works. Under choice homophily, which captures the setting when nodes preferentially accept recommendations to within group nodes,
\citep{stoica_algorithmic_2018, asikainen2020cumulative} show that such interventions lead to more exposure gains for the homophilous group which exacerbates observed homophily. In contrast, \citep{abebe_effect_2022} shows that, when the closure of triangles via friend-of-friend recommendations is not biased in favor of in-group edges, recommendations can in fact improve network integration. In our work, we investigate friend-of-friend recommendations and their impacts on network segregation and show that the impacts further depend on the length of intervention as well as the time of measurement.

\paragraph{Delayed algorithmic impacts.} Temporal dynamics associated with algorithmic interventions have been previously studied in the context of fairness in Machine Learning. These works showcase broad settings where algorithmic interventions designed to improve fairness \citep{liu2018delayed, daly2010network, akpinar2022long}, robustness \citep{milli2019social} or diversity \citep{curmei2022towards} in the short term, lead to the opposite effect in the long run. Our work uncovers similar surprising temporal dynamics in the case of link recommendations.

\section{Methodology for evaluating delayed and indirect impacts}\label{sec:methodology}
To illustrate temporal complexities of evaluating link recommendation algorithms in dynamic setting, we consider a stylized network evolution model. We propose an extension to the the classic Jackson-Rogers (JR) network evolution model \citep{jackson2007meeting}.  The JR model has been validated empirically and shown to display proprieties of real network such as decreasing diameter over time, increased edge densification and emergence of community structure \citep{jackson2007meeting, ruiz_stability_2017}. Our extension adds an optional recommendation phase to model the feedback loop between natural network evolution and algorithmic recommendations.

We analyze two simple baseline recommendation algorithms: a neighborhood-based and an affinity-based algorithm. While recommendation algorithms deployed in practice use both types of information, with potentially additional learning components, we opt for simplicity in recommendation algorithms to understand the underlying mechanisms behind evaluation phenomena in dynamic networks.
Algorithm \ref{alg:simulation} summarizes the simulation and evaluation procedure.

\subsection{Dynamic network model}\label{subsec:model}
Let $G^t = (V^t, E^t)$ be an undirected network where $V^t$ and $E^t$ are the set of nodes and edges at time $t$. Nodes are characterized by their group identity $g_i$ and latent representation $v_i$. We assume group identities are static over time and latent representations are sampled from a group-specific distribution $v_i \sim \mathcal{D}_{g_i}$. At each time step, the network evolves via new nodes which first connect "organically" the existing nodes. Further at each step, the network evolves "algorithmically" via connections mediated by a link recommender.

\paragraph{Natural growth.}
Similar to the classic JR model, upon the arrival of a new node, the network evolves in two phases: "Meeting Strangers" and "Meeting Friends". In the "Meeting Strangers" phase, the new node makes connections with existing nodes at random. Unlike the classic model where connections are made with a fixed probability, we model connections probabilities based on latent representation of the nodes $v_i$. This additional modelling assumption allows us to consider various community structure.  In the "Meeting Strangers" phase, $N_s$ candidate nodes are sampled uniformly  from the existing network. The arriving node $i$ connects with each candidate node $j$ with probability proportional to the inner product of their respective latent embeddings: $p_{i,j} = \sigma(\langle v_i, v_j \rangle)$ where $\sigma(\cdot)$ is a scaled and translated sigmoid function\footnote{We consider $\sigma(x)=\frac{1}{1+e^{-ax + b}}$ where we set $a$ and $b$ to match desired average linkage probabilities.}. 

In the subsequent "Meeting Friends" phase, the incoming node $i$ makes additional connections based on neighborhood proximity. Here $N_f$ candidate nodes are sampled from the set of nodes at distance 2 (neighbors of neighbors) from the node $i$. The node $i$ connects with each candidate node $j$ with constant probability. Optionally, we model attrition effects by considering node departures from the network with a hazard function \footnote{$h(\mathbf{a}) = cd^{\mathbf{a}} + k$ where $\mathbf{a}$ is the age of the node and $c,d,k$ are tuned to model mean and variance of lifespans in the network.} that increases with the age of the node.


\paragraph{Algorithmic intervention.}
We introduce algorithmic intervention as a third "meeting recommendations" phase, whereby nodes in the network receive link recommendation. 
This phase applies not only to incoming nodes but also to existing nodes in the network. 
Given recommendation, nodes accept them according to a behavioral model. We consider neighborhood and affinity-based recommendations. The prototypical neighborhood recommendation is the friend-of-friend (\fof{}) recommendation where candidate nodes are selected uniformly from the set of nodes at distance 2: $\mathbb{P}(\textbf{rec}_i = j) = \frac{\textbf{1}(dist(i, j) = 2}{\sum_{j'}\textbf{1}(dist(i, j') = 2}$. Conversely, affinity-based (\latent{}) recommendations utilize the latent structure rather than local neighborhood structure to make recommendations. Specifically, the \latent{} recommender computes affinity scores for all the nodes in the network in terms of inner products and recommends candidates with probability proportional to the scores: $\mathbb{P}(\textbf{rec}_i = j) \propto e^{\beta s_j}$, where $s_j = \langle v_i, v_j\rangle$ is the embedding similarity between node $i$ and node $j$. High values of softmax temperature parameter $\beta$ correspond to more deterministic recommenders. Lower values of $\beta$ lead to more random recommendations and qualitatively capture the effects of estimation noise for learning-based recommenders.

\paragraph{Behavioral models.}
We consider two behavioral models for accepting link recommendations: the constant probability baseline and the embedding-based probability where nodes accept new links based on proximity in latent space. Embedding based probabilities model more granular notions of choice homophily \citep{asikainen2020cumulative}. We consider an additional behavioral option in which upon acceptance of a new recommended link, an edge is removed at random from the set of existing edges. This option is in line with several works which model recommendations as a \emph{rewiring} \citep{asikainen2020cumulative, fabbri_exposure_2021, santos_link_2021} process that does not impact the average degree.

\begin{algorithm}[ht]
\caption{Simulating network evolution}
\SetKwInOut{Input}{input}
\Input{initial $G_0$; time-steps $T$; hazard function; 
\textit{communities:} prevalence $c_g$, latent distribution $\mathcal{D}_g$;\\
\textit{natural growth:} $N_s$, $N_f$, dist-2 connect prob $p_2$;
\textit{intervention:}  window $[\underline{t},\overline{t}]$, \texttt{recommender}, \texttt{behavior};\\
}

\For{$t$ in $1\ldots T$}{
    \textbf{Natural growth}\;
    sample incoming node $i$ according to group prevalence $c_g$ and latent distribution $\mathcal{D}_g$\;
    strangers $\gets$ samples $N_s$ nodes\;
    \For{$s$ in strangers}{
        add edge $i-s$ with probability $\propto \langle v_i, v_s \rangle$
    }
    friends $\gets$ samples $N_f$ neighbors of neighbor nodes\;
    \For{$f$ in friends}{
        add edge $i-f$ with probability $p_2$
    }
    \If{\upshape{$t\in [\underline{t},\overline{t}]$}}
    { \textbf{Algorithmic intervention}\;
    \For{node $j$ in treatment group $G^{treatment}_t$}{
        candidate = \texttt{recommender}($j, G_t$)\;
        \If{\upshape{\texttt{behavior}($j$, candidate) = accept}}{ add edge $j-$candidate}
    }
    }
    nodes to remove $\gets$ hazard function
}
\end{algorithm}\label{alg:simulation}

\subsection{Structural metrics}\label{subsec:metrics}


\emph{Clustering coefficient}: Clustering coefficient of a node is the ratio of triangles it forms $\Delta_i$ and the maximum number of triangles it could have formed given it's current degree $d_i$: $c_i = \frac{2\Delta_i}{d_i(d_i-1)}$. The clustering coefficient captures the micro-level cohesion in the network \citep{ferrara2022link}. The average clustering coefficient is the metric averaged over all the nodes. Averaging the metric over communities measure clustering at the community level.

\emph{Gini coefficient}: We measure the inequality in the degree distribution via Gini coefficient. An increase of the Gini coefficient resulting from the use of recommendations is often used to demonstrate biases in link recommendation \citep{ferrara2022link, fabbri_exposure_2021, fabbri2022exposure}. The Gini coefficient can be computed for an ordered list of node degrees as: $G = \frac{2\sum_{i=1}^n id_i}{n \sum_{i=1}^n d_i} - \frac{n+1}{n}$. Similarly, this metric can be computed for the entire graph or restricted to communities.

\emph{Homophily}: We define a monocromatic and bichromatic edges to be an edges that links two nodes from the same community or nodes from different communities, respectively. The homophily of a community measures the propensity of nodes to favor within group connections compared to a non-preferential baseline. We compute the homophily of a community $g$ as follows: $H_g = \frac{\vert E_{gg}\vert}{\vert E_g\vert} - \frac{n_g}{n}$, where $\vert E_{gg}\vert$ denotes the number of monochromatic edges within $g$, $\vert E_{g}\vert$ denotes the number of total edges that have at least one node in community $g$. Finally  $n_g$ denotes the size of $g$, and $n$ denotes the total size of the network, the ratio $\frac{n_g}{n}$ is the baseline fraction of within-group links when nodes have no group-based preferences. The vast majority of existing simulation based evaluation study changes in homophily or the closely related notions of within-community and between-community edges \citep{fabbri_effect_2020, abebe_effect_2022, daly2010network}.

\subsection{Evaluation}\label{subsec:evaluation}

We denote by $\mathcal{G}$ the evolution of the network under natural dynamics
and by $\mathcal{G}(\texttt{rec},[\underline{t}, \overline{t}])$ the evolution 
when the network receives link recommendation according to 
$\texttt{rec}$ algorithm over the $[\underline{t}, \overline{t}]$ intervention interval. The subscript $T$ in $\mathcal{G}_T(\cdot)$ is used to denote the snapshot of the network at time $T$. For a structural metric $m$, we denote by $m(\mathcal{G}_t(\texttt{rec},[\underline{t}, \overline{t}]))$ the value of this metric evaluated at time $t$. Given a network evolution model, we simulate the counter-factual trajectories of the metrics for intervention intervals  $[\underline{t}, \overline{t}]$ of different lengths.

\paragraph{Total effects.}
The total effect of intervening in a network evolution $\mathcal{G}$ with recommender \texttt{rec} for intervention interval $[\underline{t}, \overline{t}]$ on the metric $m$ at time $T$ is defined as the difference between two counterfactual measurements:
\begin{align*}
    \text{Effect}_T(\mathcal{G},\texttt{rec},[\underline{t}, \overline{t}]) := m(\mathcal{G}_T(\texttt{rec},[\underline{t}, \overline{t}])) - m(\mathcal{G}_T).
\end{align*}
This definition compares between the ``treatment'' universe where the whole network received an algorithmic intervention and a counterfactual ``control'' universe where the network evolved without algorithmic recommendations.

\paragraph{Delayed effects.}
When the measurement occurs during the intervention interval, i.e. $T \in [\underline{t}, \overline{t}]$ we call the effect \emph{immediate}. Conversely, when the measurement occurs after the intervention stops, i.e. $T > \overline{t}$, we can define the notion of \emph{delayed} impact as: 
$$\text{DelayedEffect}_{T}(\mathcal{G},\texttt{rec},[\underline{t}, \overline{t}]) = \text{Effect}_{T}(\mathcal{G},\texttt{rec},[\underline{t}, \overline{t}]) - \text{Effect}_{\overline{t}}(\mathcal{G},\texttt{rec},[\underline{t}, \overline{t}]) $$

The notion of delayed impacts allows us to characterize the impact of link recommendations into three broad categories: diminishing, amplifying and persistent based on the sign of $\text{DelayedEffect}_{T}(\mathcal{G},\texttt{rec},[\underline{t}, \overline{t}])$. Note that the delayed impacts measure the difference in effect sizes between time $\overline{t}$ and time $T$, rather than the difference in the metric $m(\mathcal{G}_T(\texttt{rec},[\underline{t}, \overline{t}]))-m(\mathcal{G}_{\overline{t}}(\texttt{rec},[\underline{t}, \overline{t}]))$. This two notions are equivalent only when the metric remains constant from time $\overline{t}$ to time $T$ under natural network evolution dynamics.

\paragraph{Indirect effects.}\label{subsec:indirect_methodology}

\begin{figure}
    \centering
    \includegraphics[trim={5cm 5.5cm 8cm 9cm}, clip, width=0.9\linewidth]{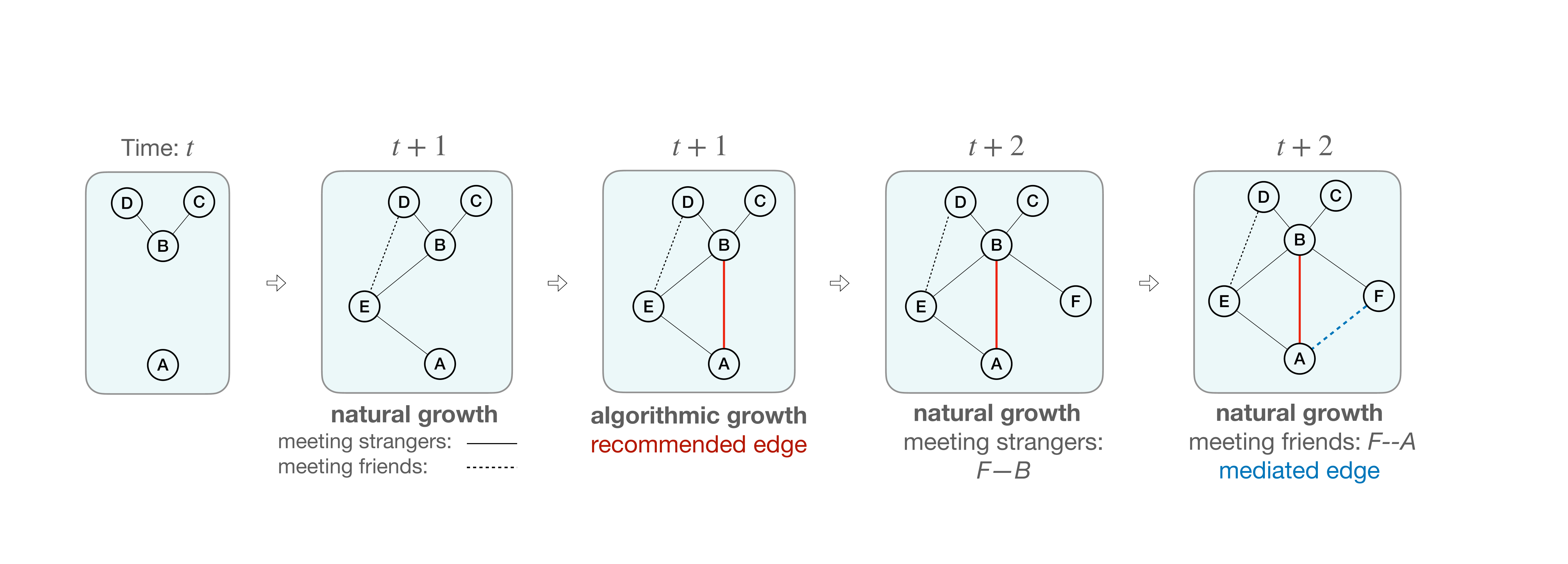}
    \caption{Indirect effects of recommendation though mediated edges: Edge $A-B$ is a recommended edge and contributes to the direct effect of the recommendation. Edge $F-A$ is a mediated edge as node $A$ would not be reachable from $F$ in the absence of $A-B$.
    } 
    \label{fig:direct_indirect}
\end{figure}

The temporal evolution of the network in the presence of link recommendations is affected \emph{directly} by the addition of algorithmic edges; but also \emph{indirectly} as the addition of algorithmic edges biases the natural growth dynamics. Recommendations have an indirect impact on natural dynamics in the "Meeting Friends" phase of the network evolution model. In this phase, the arriving nodes connect with nodes at distance 2 (neighbors of neighbors). In the presence of algorithmic edges, there can be nodes at distance 2 that are only reachable via algorithmic edges. If the incoming node connects to such a node, the resulting edge is \emph{mediated} by recommendation. Figure \ref{fig:direct_indirect} illustrates the mechanism through which recommendations mediate the organic creation of new edges.



To measure indirect impacts, we design a counterfactual experimental procedure to remove the influence of algorithmic edges on natural growth, and thus discount mediated edges. The counterfactual procedure is defined as follows: upon arrival of a new node $i$, we modify the "Meeting Friends" phase to only consider node candidates that are at distance 2 from $i$ via non-algorithmic edges. For this counterfactual, the network contains no mediated edges, and therefore it isolates the \emph{direct} effects. Formally, we refer to this unmediated evolution as $\tilde{\mathcal{G}}(\texttt{rec},[\underline{t}, \overline{t}])$.  
Finally we can define direct effects as the difference between the metric evaluated for the unmediated counterfactual and the natural growth trajectory. The indirect effects are the difference between total effect and direct effect:
\begin{align*}
\text{DirectEffect}_T(\mathcal{G},\texttt{rec},[\underline{t}, \overline{t}]) &= m(\tilde{\mathcal{G}}_T(\texttt{rec},[\underline{t}, \overline{t}])) - m(\mathcal{G}_T),\\
\text{IndirectEffect}_T(\mathcal{G},\texttt{rec},[\underline{t}, \overline{t}]) &=  \text{Effect}_T(\mathcal{G},\texttt{rec},[\underline{t}, \overline{t}])  - \text{DirectEffect}_T(\mathcal{G},\texttt{rec},[\underline{t}, \overline{t}]).
\end{align*}

\paragraph{Longitudinal evaluation.}
In a simulated setting, one has access to both counterfactuals: the evolution of the network in the presence of recommendations and in their absence. In longitudinal evaluation which are commonly used in observational studies, effects are measured as the difference in a metric before and after the intervention:
$$\widehat{\text{Effect}}^{obs}_T(\mathcal{G},\texttt{rec},[\underline{t}, \overline{t}]):= m(\mathcal{G}_T, \texttt{rec},[\underline{t}, \overline{t}]) - m(\mathcal{G}_{\underline{t}})$$
This measure is biased whenever  $ m(\mathcal{G}_{T}) \ne m(\mathcal{G}_{\underline{t}}) $.

\paragraph{A/B evaluation.}
The lack of valid counterfactuals motivates the use of A/B testings in settings with interventional access. In an A/B test, nodes are divided into two groups: treatment and control; the treatment group receives recommendations, while the control group does not. However, in networks, the assumptions of Stable Unit Treatment Value Assumption (SUTVA) \citep{imbens_causal_2015} do not hold due to network interference. There are various methods for choosing treatment nodes, as well as methods to correct for network interference in estimation procedures.

Let a scheme for choosing treatment nodes exist and let $\mathcal{G}^{AB}(\texttt{rec}, [\underline{t}], \overline{t})$ be the network evolution where a group of nodes are assigned to the treatment group and receive recommendations. To estimate the impact of recommendations on a metric of interest, we compare the metric's values on the treatment and control groups:

$$\widehat{\text{Effect}}^{AB}_T(\mathcal{G},\texttt{rec},[\underline{t}, \overline{t}]):= m^{treatment}(\mathcal{G}^{AB}_T, \texttt{rec},[\underline{t}, \overline{t}]) -  m^{control}(\mathcal{G}^{AB}_T, \texttt{rec},[\underline{t}, \overline{t}]). $$

In evaluating the impact of recommendations, we consider both naive estimators that do not correct for network interference, and more sophisticated ones that take externalities into account.

\section{Counterfactual evaluation}\label{sec:experiments}
We simulate counterfactual scenarios by applying different recommendation interventions for various time periods. Although accessing counterfactuals in reality is infeasible, these simulations offer insights into dynamic and temporal phenomena and provide grounded foundation for subsequent sections.
\subsection{Setup}
The results in sections \ref{subsec:delayed} and \ref{subsec:indirect} consider a simple experimental setup to illustrate delayed and indirect effects of \fof{} and \latent{}\footnote{For \latent{} recommendations we use softmax temperature $\beta = 10$} recommenders. In the baseline setup we consider two equally sized communities. We sample latent embeddings for the nodes independently from $\mathcal{N}(\mu_c, \sigma I)$, with $\mu_1 = [0,1]$ and $\mu_2 = [1, 0]$; for both communities the variance of the embeddings is set to $\sigma = 0.05$. We sample 50 nodes in each group and initialize edges by connecting pairs of nodes $i-j$ with probability proportional to the inner product of their embeddings. Then, for each node in the network, we consider their neighbors at distance 2 and connect to them with probability $p_1=0.05$. This results in a slightly homophilic initial network with homophily $h1=h2\approx 0.1$.  Upon initialization, at each time step $5$, new nodes arrive. For natural growth we consider $N_s = N_f = 100$ and the connection probability of connecting to a candidate node in the "Meeting Friends" phase: $p_2= 0.05$. Further, we assume a behavioral model where nodes accept recommended edges with a constant probability, $0.5$.
At each time step we measure structural metrics of the network such as clustering coefficient, Gini coefficient and homophily. We repeat each trajectory for 5 random seeds and report average results along with confidence bands.
\footnote{Code for reproducing experimental results can be found \href{https://anonymous.4open.science/r/delayed-indirect-CB60/README.md}{here}.}

In section \ref{sec:group_structure} we vary these assumptions by considering majority-minority structure, differentiated homophily, and within group heterogeneity. Finally in Appendix \ref{app:rewire_acc} and \ref{app:rec_variants}, we consider additional behavioral assumptions, different variants of the underlying dynamics and modifications to the recommendation algorithms.



\subsection{Delayed effects}\label{subsec:delayed}
\begin{figure}
    \centering
    \includegraphics[width=0.9\linewidth]{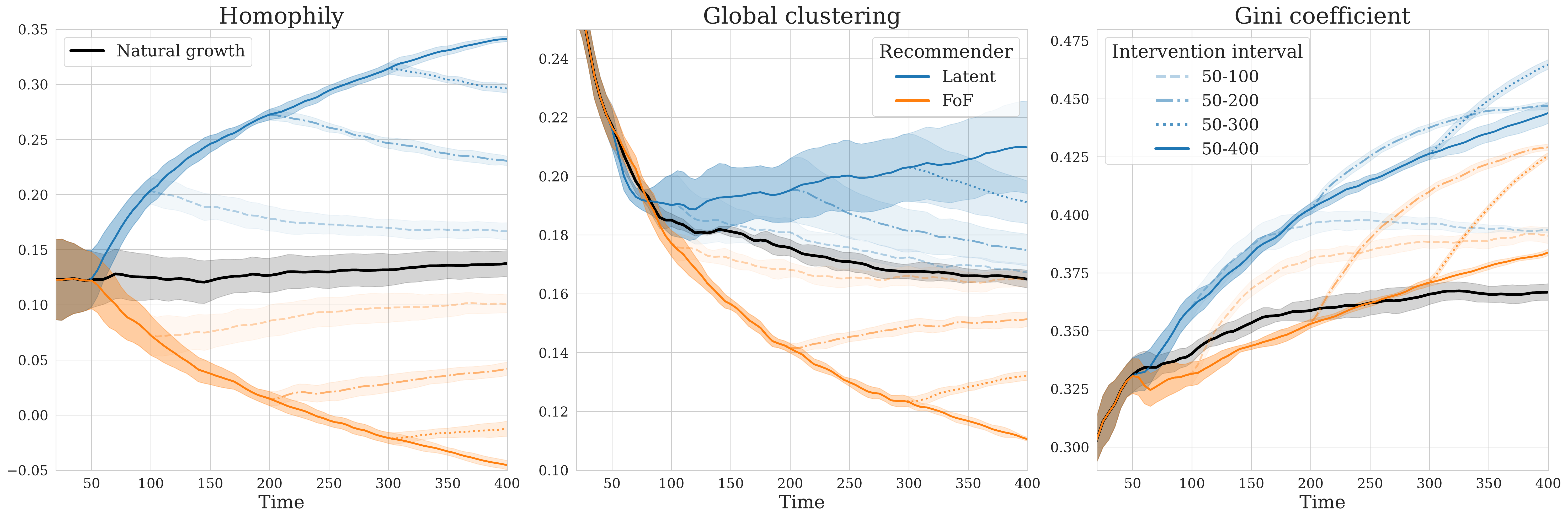}
    \caption{Trajectory of homophily, global clustering and Gini coefficient for \latent{} and \fof{} recommendations applied over varying intervention intervals. The solid black line represents the evolution of the metric under natural growth dynamics. The blue lines represent trajectories for the \latent{} recommender whereas the orange lines correspond to \fof.  The shaded area corresponds to the 95\% confidence intervals calculated over 5 independent trajectories.}
    \label{fig:gini_delayed}
\end{figure}

We find that affinity and neighborhood based recommenders have opposite long term effects on homophily and clustering coefficients. \latent{} leads to increases in homophily as well as global clustering whereas \fof{} recommendations decrease them. Both recommenders have diminishing delayed impacts with respect to homophily and clustering as upon end of the intervention the trajectory of the metric regresses to the counterfactual natural growth trajectory. Meanwhile, effects with respect to the Gini coefficient are qualitatively different. \latent{} recommendations lead to increases in degree inequality both in the short and in the long term. Conversely, \fof{} recommendations decrease the Gini coefficient in the short term but increase it in the long term. Figure \ref{fig:gini_delayed} illustrates these findings.

\paragraph{Clustering and homophily.}
In the case of \latent{} recommendations, if nodes $i$, $j$, and $k$ are similar in embedding space, it is likely they are from the same community and all three pairs of edges have been proposed as recommendations, leading to a strong bias for closing triangles within the community. This results in increased homophily and clustering, especially with the behavioral assumption that nodes accept links based on embedding similarity (see Fig. \ref{fig:bevarior_variants_c, fig:bevarior_variants_d}). The bias is further exacerbated for high-temperature $\beta$. Conversely, with a low $\beta$ value, which corresponds to a more stochastic recommender, it is less likely that all three edges of a given triplet will close, thus slightly lowering the homophily value compared to the high-$\beta$ case.

The \fof{} recommender intervention mimics the "Meeting Friends" phase of natural growth, which has a bias towards forming cross-community links, resulting in decreased homophily. Decreased clustering occurs because random \fof{} recommenders lack the bias of connecting nodes with a large common neighborhood, unlike neighborhood-based models such as Adamic-Adar \citep{adamic2003friends}, which favor links with nodes with a large number of common neighborhoods and thus increase clustering (see Fig. \ref{fig:rec_variants}).


\paragraph{Gini coefficient.}
The Gini coefficient for recommendations shows amplifying delayed effects. The natural growth trend exhibits a slight increase in inequality over time. \latent{} recommendations exacerbate the wealth gap between popular and unpopular nodes through a bias towards nodes with high embedding norms, creating the "rich-get-richer" effect. \fof{} recommendations initially reduce degree inequality as they are less biased towards popular nodes.
Surprisingly, once recommendations stop, the Gini coefficient increases dramatically for both recommenders, particularly for \fof{}. This is due to a "relative-rich-minority" effect caused by differences in edge density between 'older' and 'newer' nodes. Over time, as natural growth is continuing to take place, the set of "rich" nodes becomes relatively smaller, resulting in higher inequality as measured by the Gini coefficient. If a rewiring behavioral model is assumed  (see Fig. \ref{fig:bevarior_variants_b, fig:bevarior_variants_d}), where nodes receiving recommendations do not change their degree, the delayed impacts diminish, similar to the case of clustering and homophily.

\subsection{Indirect effects}\label{subsec:indirect}
The "Meeting Friends" phase occurs in feedback loop with algorithmic recommendations, leading to the mediation phenomenon illustrated in Figure \ref{fig:direct_indirect}. We find that mediated edges are common, long-lasting and biased. For instance, in the case of \latent{} recommendations, the mediated edges are more likely to connect nodes from the same community, resulting in significant effects on homophily. 


\begin{figure}
    \centering
    \includegraphics[width = 0.9\linewidth]{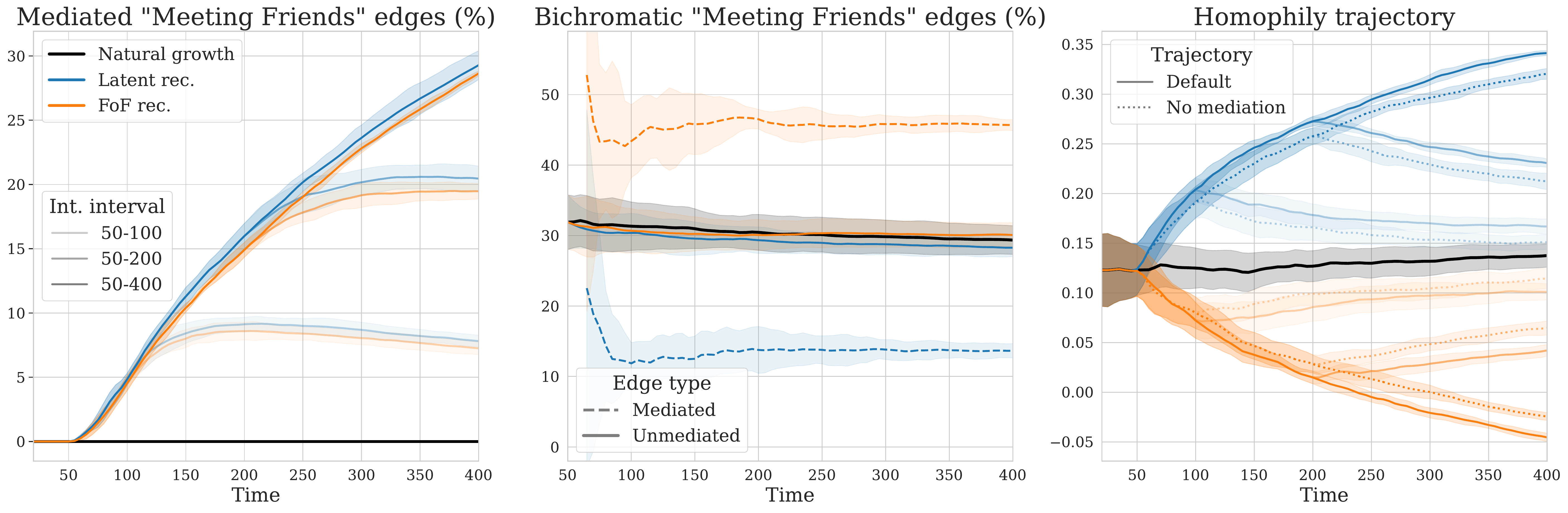}
    \caption{Indirect effects: Solid black line corresponds to natural growth dynamics. The blue and orange lines represent trajectories for \latent{} and \fof{} recommenders, respectively. Shaded areas corresponds to 95\% confidence bands.  Left plot illustrates the prevalence of algorithmically mediated edges in the "Meeting Friends" phase of the natural growth dynamics. Dashed and dotted lines correspond to various intervention intervals. Center plot illustrates the bias in the mediated edges relative to unmediated ones. Solid lines correspond to fraction of bicromatic edges among unmediated edges created in the "Meeting Friends" phase; the dashed line records the fraction of bicromatic mediated edges. Right plot shows the trajectory of homophily metric. Dotted lines correspond to the trajectory of the metric for the unmediated counterfactual evolution of the network.
    }
    \label{fig:prevalence_persistance_bias}
\end{figure}
\paragraph{Prevalence, persistence and bias of mediated edges.}
 Mediated edges play a substantial role in the "Meeting Friends" stage of natural growth dynamics. Figure \ref{fig:prevalence_persistance_bias} shows that the proportion of mediated edges increases with the duration of the intervention period and remains constant even after recommendations have stopped for both \latent{} and \fof{} recommenders. The bias of mediated edges in terms of bichromaticism is distinct for each recommender, relative to the proportion of bichromatic edges under natural growth.
For instance, about a third of unmediated "Meeting Friends" edges are bichromatic for both recommenders. However, the proportion of bichromatic edges among mediated edges is lower for \latent{} and higher for \fof{}. Our findings about homophily in Section \ref{subsec:delayed} suggest that algorithmic edges are biased towards monochromatic for \latent{} and bichromatic for \fof{}. This experiment supports the hypothesis that recommendations have a compounding effect by inducing biases in the natural network evolution.
 
 \paragraph{Counterfactual measurements of indirect effects.}

To measure the indirect effects on structural metrics, we apply the counterfactural procedure from Section \ref{subsec:indirect_methodology} and isolate the direct effects.  We find that without mediation, structural metrics trend faster to their natural evolution. Figure \ref{fig:prevalence_persistance_bias} shows the comparison of homophily under natural growth, algorithmic growth, and unmediated algorithmic growth. The difference between the latter two quantifies the magnitude of indirect effects. Additionally, indirect effects grow relatively stronger over time, even after recommendations end, highlighting the persistence of indirect impacts.

\subsection{Impact of group structure}\label{sec:group_structure}
Previous studies have highlighted the unequal impact of recommendations on different communities, especially when they are divided into majority and minority groups and display varying levels of homophily \citep{stoica_algorithmic_2018, ferrara2022link, fabbri_effect_2020}. Here, we examine the effects of differential homophily between majority and minority groups as well as within-community heterogeneity.
majority and minority group and the role of within-community heterogeneity. 
 
\begin{figure}
     \centering
     \includegraphics[width = 0.9\linewidth]{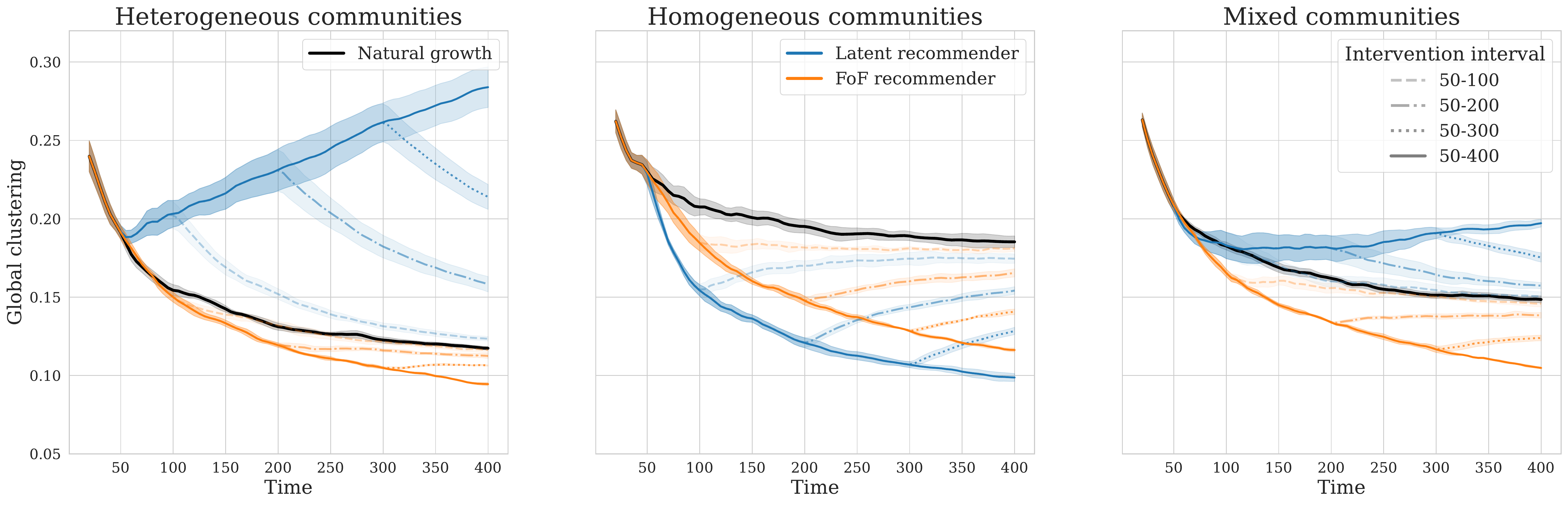}
     \caption{Effects of within-group heterogeity. Black trajectory  corresponds to the natural evolution of the global clustering metric. Blue and orange line indicate the trajectory for \latent{} and \fof{} recommender. Left plot: two heterogenous groups. Center plot: two homogenous group. Right plot: one homogenous and one heterogeneous group. }
     \label{fig:homo_clustering}
 \end{figure}

 \paragraph{Community heterogeneity.}
We examine the impacts of homogeneity and heterogeneity in the latent representation of nodes on the results of link recommendation. We model within-group heterogeneity by varying the variance of the latent embeddings. In the heterogeneous setting, we set the variance of the embedding distribution to $\sigma^2 = 0.1$, and to $\sigma^2 = 0.01$ in the homogeneous setting; in the extreme case when the variance is 0 this recovers the JR variant of \citep{abebe_effect_2022}. 
Figure \ref{fig:homo_clustering} shows that when there is high within-group heterogeneity, \latent{} recommendations greatly increase the global clustering coefficient, while  for high within-group homogeneity, \latent{} recommenders have the opposite effect, reducing global clustering. This unexpected phenomenon is due to \latent{} recommendations favoring nodes with high embedding norms. For any existing edge $i-j$, there is a disproportionately high chance that both $i$ and $j$ will receive recommendations concentrated on a small subset of large-normed nodes, creating closed triangles. In the homogeneous setting, such 'collisions' are less likely to occur, as the probability of two nodes independently being recommended with the same node is lower. These findings highlight the importance of investigating not only between-group differences but also within-group differences.

\begin{figure}
    \centering
    \includegraphics[width = \linewidth]{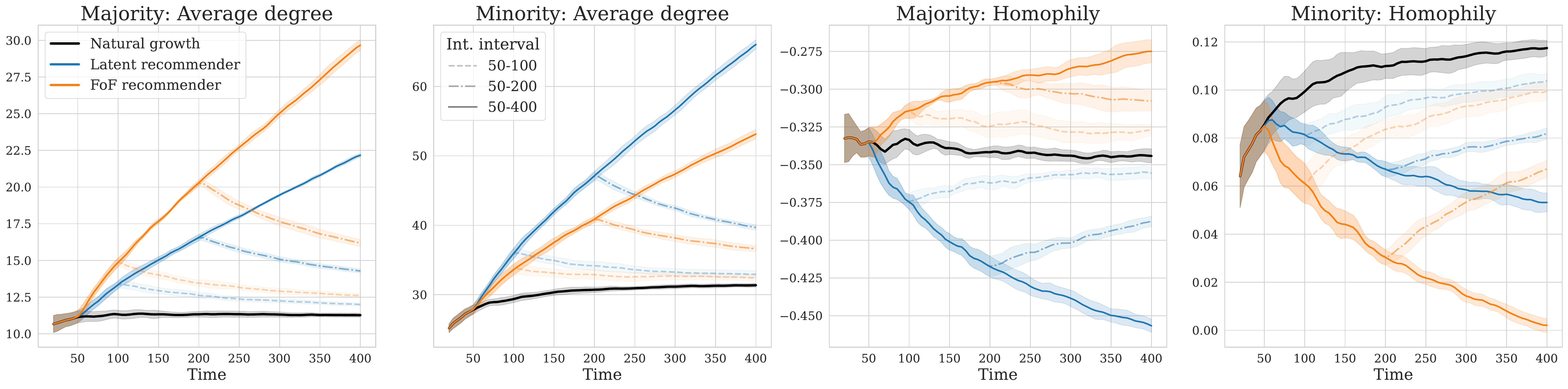}
    \caption{Effect of homophily: Trajectory of average degree and community homophily for \latent{} and \fof{} recommenders for the heterophilic majority and homophilic minority. }
    \label{fig:minority_majority}
\end{figure}

 \paragraph{Homophilic minority and heterophilic majority.}
In previous studies, it has been observed that homophilic minority groups receive an excessive amount of exposure from recommendations, leading to increased disparities in homophily between groups. We simulate the network evolution for a majority fraction of $60\%$, $\mu_1 = [0,1]$ and $\mu_2 = [1.2, 1]$ and find that our results support this conclusion for \latent{} recommendations. In contrast, Figure \ref{fig:minority_majority} shows that while \fof{} increases homophily for the majority, it decreases it significantly for the minority. The \fof{} recommender amplifies the visibility of the majority more than that of the minority. This phenomenon can be explained by the heterophilic nature of the majority group, where most of its direct connections are with minority nodes. Though the minority group is homophilic, there is a larger probability that a majority node $i$ is recommended with majority nodes at distance 2, because $i$'s neighbors' neighbors that are in the minority group are more likely to also be direct neighbors of $i$.






\subsection{Evaluation Biases.}
As we do not have access to the full range of counterfactual measurements, we often resort to either longitudinal or A/B evaluations. However, these methods have their limitations and may fail to accurately capture the impacts of recommendations. In this section, we explore the potential biases and limitations of these evaluation procedures.

 \paragraph{Longitudinal evaluation.}
Simulating a longitudinal evaluation mimics an observational study. In this setting, a single trajectory is observed. The estimated total effect of link recommendations is the temporal difference between the value of the metric before and after the intervention. If the underlying dynamics for a metric are stationary, as is the case for homophily in Figure \ref{fig:gini_delayed}, then a longitudinal evaluation is unbiased. Conversely, when under natural dynamics a metric is non-stationary, a naive comparison between the initial network at time $\underline{t}$ and the network at time $T$ could yield qualitatively misleading estimates. For instance, in Figure \ref{fig:gini_delayed}, when measuring the impact of \fof{} on clustering coefficient, a longitudinal measurement would compare the solid orange trajectory between time $\underline{t}=50$ and $T=400$, thus over-estimating about double the size of the true effects. Furthermore, the naive observational measurement for \latent{} would incorrectly suggest a negative effect on network clustering, when in reality it has a positive impact compared to the natural evolution. 

\begin{figure}
    \centering
    \begin{subfigure}[b]{\linewidth}
    \centering
    \includegraphics[width=\linewidth]{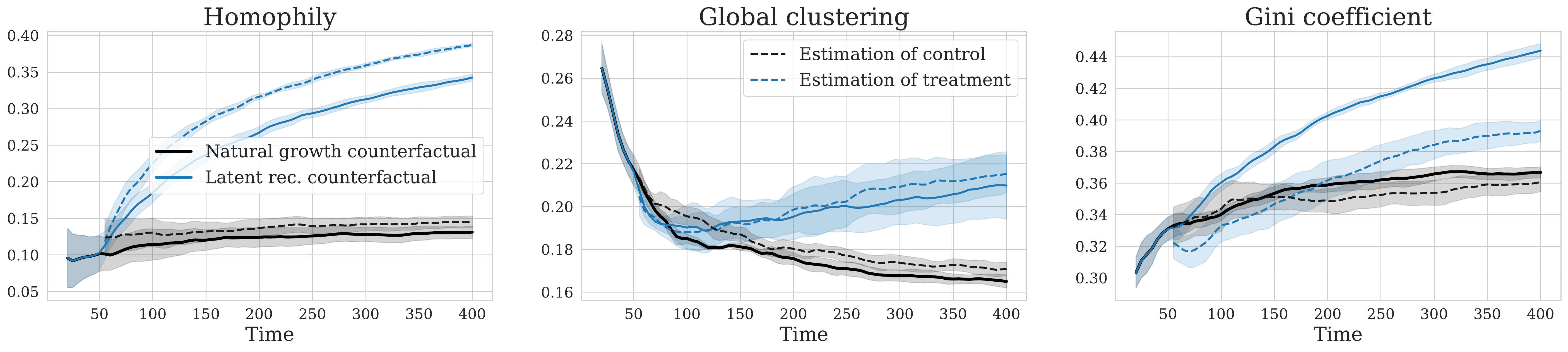}
    \caption{\latent{} recommender}
    \label{fig:ab_latent}
    \end{subfigure}

    \begin{subfigure}[b]{\linewidth}
    \centering
    \includegraphics[width=\linewidth]{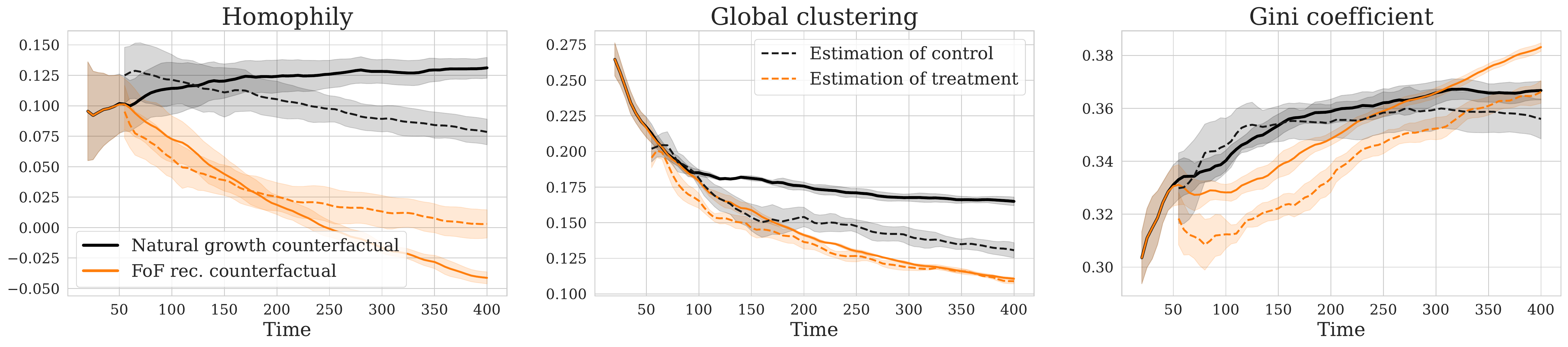}
    \caption{\fof{} recommender}
    \label{fig:ab_fof}
    \end{subfigure}
    \caption{A/B Evaluations: Solid lines correspond to ground truth counterfactual evaluation of homophily, clustering and Gini coefficient. Dashed lines correspond to trajectories estimated from running an A/B test.}
    \label{fig:ab}
\end{figure}

 \paragraph{A/B evaluation.}
A/B evaluations in dynamic networks become complex due to changes of network structure over time caused by natural dynamics and algorithmic interventions. We simulate A/B tests aimed at estimating causal effects of \latent{} and \fof{} link recommendations. We perform random treatment ($p=0.5$) on node assignment for clustering and Gini coefficient and community-based treatment assignment for measuring homophily. 
The estimates in Figure~\ref{fig:ab} are adjusted for network interference. Detailed computation of the adjusted estimates as well as the corresponding plots for the naive cases can be found in Appendix \ref{app:ab}. 
The quality of A/B estimates varies based on the metric and intervention. For example, A/B tests overestimate the impact on homophily for \latent{} but underestimate it for \fof{}. The estimation for clustering is accurate for \latent{} but greatly underestimated for \fof{}. For the Gini coefficient, both \latent{} and \fof{} interventions underestimate the metric for the treatment group while accurately estimating the metric's evolution under natural growth for the control group. Accross most settings, the quality of the metric deteriorates over time, further supporting the existence of dynamic effects that compound network interference effects.


\section{Discussion and Future work}\label{sec:discussion}
In this study, we explored the dynamic effects of link recommendations on network evolution through simulations. Our proposed extension of the Jackson-Rogers model provides insight into the impact of link recommendations on network structure. Emphasizing the importance of temporal dynamics and measurement timing, our simulations revealed surprising and persistent effects of link recommendations on network structure. 

Using synthetic data and simple network evolution models and recommendation algorithms, we answered "what-if" scenarios in a controlled setting, providing a valuable first step in understanding these effects in real networks. Our results showed that link recommendations can have delayed and indirect impacts on network structure, with long-lasting effects even after recommendations have ceased, which result in significant cascading indirect effects over time. This highlights the need for further research in evaluating link recommendation algorithms in dynamic networks.

Finally, our study sheds light on the potential biases that can arise in evaluating link recommendation algorithms in dynamic networks. We find that evaluating metrics longitudinally or using A/B tests can result in biased estimates when the underlying network dynamics are not stationary. This highlights the need for advanced estimation procedures that consider both network interference and underlying dynamic effects to accurately assess the impacts of link recommendation algorithms.

We identify several opportunities for future research. To improve the validity of our conclusions, one avenue is to validate modeling assumptions against real-world networks. Refining the modeling assumptions, such as allowing for non-recommender-driven edge creation between existing nodes, could better reflect reality and potentially result in greater indirect effects. Another important area of interest is to further develop methods for measuring direct and indirect effects in different network formation models.

A second direction of future work is to study the downstream impacts of link recommendation. Modeling the node embeddings as a dynamic property, influenced by the local neighborhood through biased assimilation \citep{dandekar_biased_2013} or mere exposure effects \citep{curmei2022towards}, could provide insight into how opinion formation on networks is influenced by recommendations. Finally, examining recommendation scenarios where network edges in the social network are formed by content recommendations, rather than user recommendations, which is typical of social media sites like Instagram,  is a promising area for investigation.
\newpage

\bibliographystyle{plainnat}
\bibliography{main}  

\appendix

\newpage
\section{Additional Experiments}

\subsection{Behavioral assumptions} \label{app:rewire_acc}
\begin{figure}
    \centering
    
    \begin{subfigure}[b]{0.85\linewidth}
    \centering
    \includegraphics[width=\linewidth]{figures/delayed_effects.pdf}
    \caption{Edge addition + Constant recommendation acceptance probability (plot in Sec. \ref{subsec:delayed})}
    \label{fig:bevarior_variants_a}
    \end{subfigure}

    \begin{subfigure}[b]{0.85\linewidth}
    \centering
    \includegraphics[width=\linewidth]{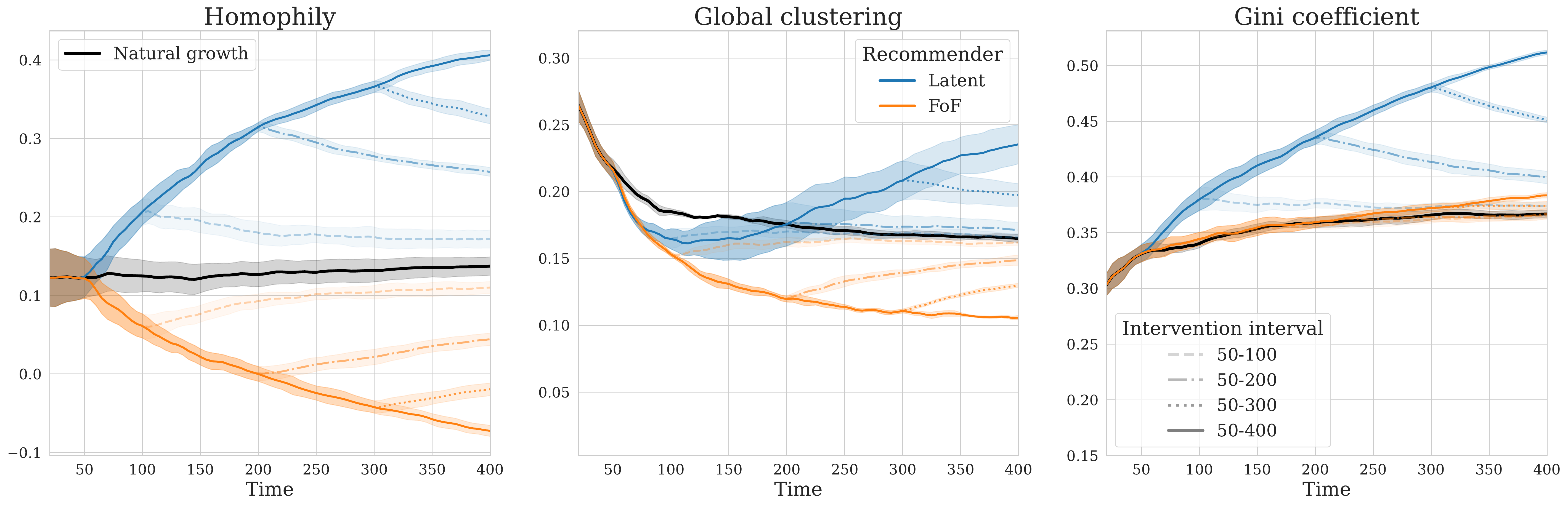}
    \caption{Edge rewiring + Constant recommendation acceptance probability}
    \label{fig:bevarior_variants_b}
    \end{subfigure}

    \begin{subfigure}[b]{0.85\linewidth}
    \centering
    \includegraphics[width=\linewidth]{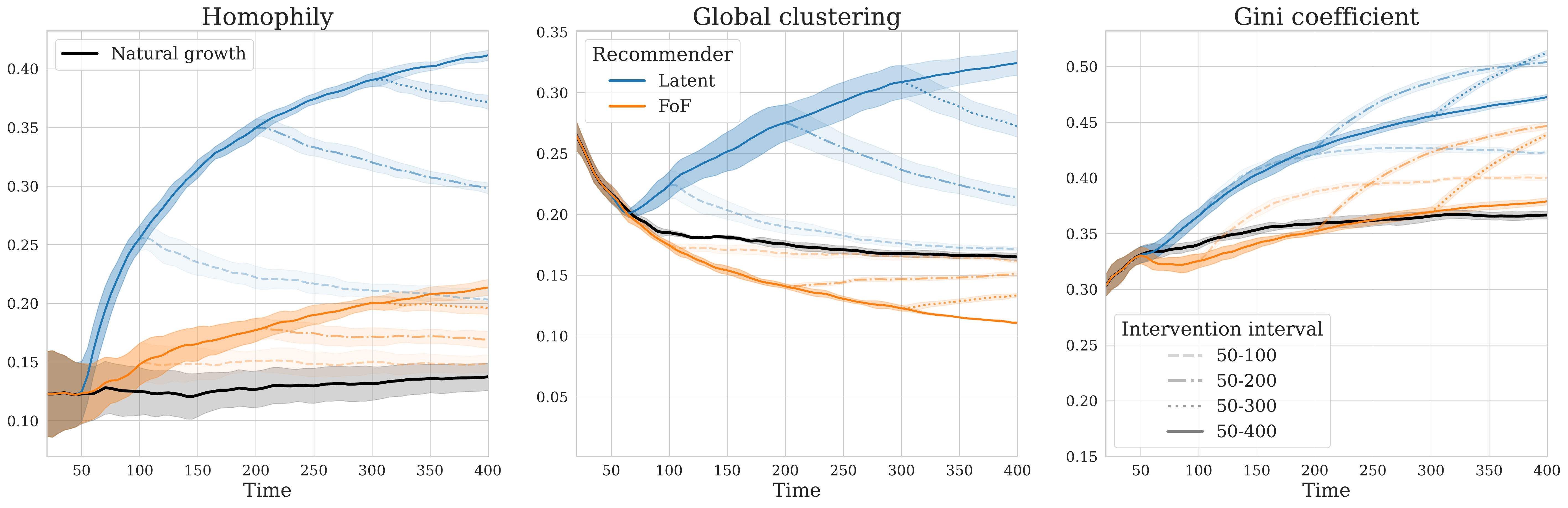}
    \caption{Edge addition + Choice homophily}
    \label{fig:bevarior_variants_c}
    \end{subfigure}

        \begin{subfigure}[b]{0.85\linewidth}
    \centering
    \includegraphics[width=\linewidth]{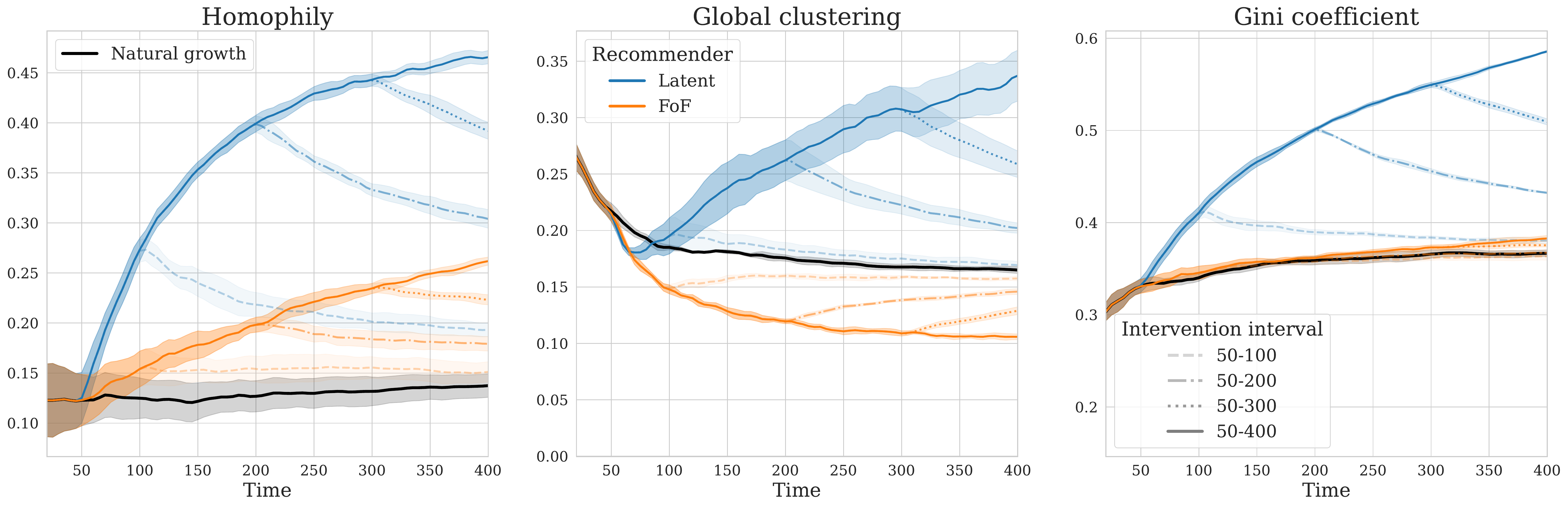}
    \caption{Edge rewiring + Choice homophily}
    \label{fig:bevarior_variants_d}
    \end{subfigure}
    
    \caption{Counterfactual trajectory of homophily, global clustering and Gini coefficient for \latent{} and \fof{} recommendations applied over varying intervention intervals. The solid black line represents the evolution of the metric under natural growth dynamics. The blue lines represent trajectories for the \latent{} recommender whereas the orange lines correspond to \fof.  The shaded area corresponds to the 95\% confidence intervals calculated over 5 independent trajectories. }
    \label{fig:bevarior_variants}
\end{figure}

In this section we investigate two behavioral assumptions: edge rewiring and choice homophily. Under edge rewiring assumption upon accepting a recommended edge the node will remove at random one of it's existing edges. Under choice homophily nodes accept recommendations according to embedding similarity rather than with a constant probability, assumed in the main body of the paper. Figure \ref{fig:bevarior_variants} shows the trajectories of homophily, clustering and Gini coefficient under different behavioral assumption. In the case of edge-rewiring (Fig. \ref{fig:bevarior_variants_b}, \ref{fig:bevarior_variants_d}) the delayed impacts to Gini coefficient diminish. The behavioral assumption that nodes accept preferentially nodes with high embedding similarity unsurprisingly leads to further increases in observed homophily for both \latent{} and \fof{} recommenders (Fig \ref{fig:bevarior_variants_c}, \ref{fig:bevarior_variants_d}).

\subsection{Intervention Variants} \label{app:rec_variants}
\begin{figure}
    \centering
    \begin{subfigure}[b]{0.9\linewidth}
    \centering
    \includegraphics[width=\linewidth]{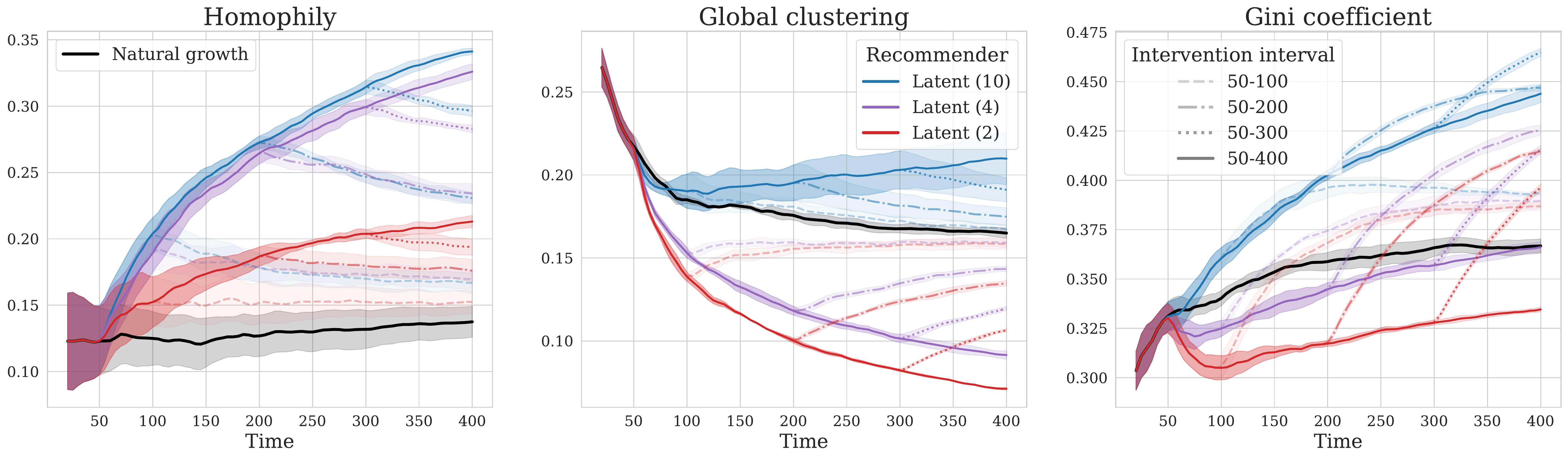}
    \caption{\latent{} recommender}
    \label{fig:rec_variants_a}
    \end{subfigure}

    \begin{subfigure}[b]{0.9\linewidth}
    \centering
    \includegraphics[width=\linewidth]{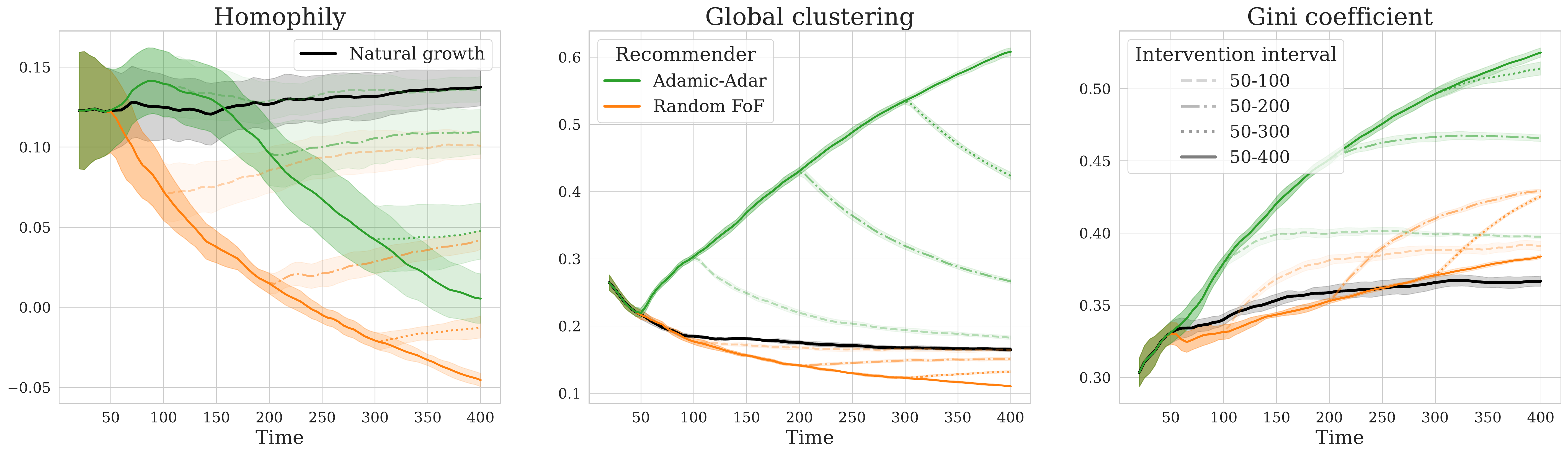}
    \caption{\fof{} recommender}
    \label{fig:rec_variants_b}
    \end{subfigure}
    \caption{Counterfactual trajectories for variants of \latent{} and \fof{} recommenders}
    \label{fig:rec_variants}
\end{figure}

In Fig. \ref{fig:rec_variants_a} we consider different choices of temperature parameter $\beta$ for the \latent{} recommender. All recorded metrics increase as the recommendations become more deterministic (uncreasing $\beta$). Note that $\beta =4$ and $\beta=10$ are nearly indistinguishable in the homophily graph whereas $\beta=4$ and $\beta=2$ are close in the graph of the clustering coefficient. This suggests that different metrics have different sensitivities to changes in recommendation intervention.

In Fig. \ref{fig:rec_variants_b} we compare \fof{} with a neighborhood based recommendation algorithm that recommends node pairs with highers Adamic-Adar index. The Adamic-Adar index assigns a higher similarity score to pairs of low degree nodes that share many neighbors in common. This change leads to extreme increase in the clustering coefficient.

\subsection{Naive and Interference Adjusted estimates for A/B evaluation}\label{app:ab}
First we comment on the naive and adjuted estimation procedure for each metric.
\paragraph{Homophily.} Naively, we estimate homophily for an A/B evaluation by assigning one community as treatment and the other as control and measuring the homophily for each community. To adjust the estimate for the control group, we discount the algorithmic edges received by the control group, whereas to adjust for the treatment, we double the count of the algorithmic edges that the treatment group forms with the control group.
\paragraph{Clustering.} Naively, we estimate the group-level clustering coefficient for treatment and control by averaging node-level clustering for treatment and control groups respectively. We adjust the estimate for the treatment/control group by calculating its clustering coefficient only among the treatment/control neighbors, i.e., in a sub-network with only the treatment/control nodes.
\paragraph{Gini coefficient.} We naively estimate Gini coefficient by simply measuring the value within the treatment and control groups, respectively. We adjust Gini coefficient by discounting the algorithmic edges for the control group.

\begin{figure}
    \centering
    \begin{subfigure}[b]{0.9\linewidth}
    \centering
    \includegraphics[width=\linewidth]{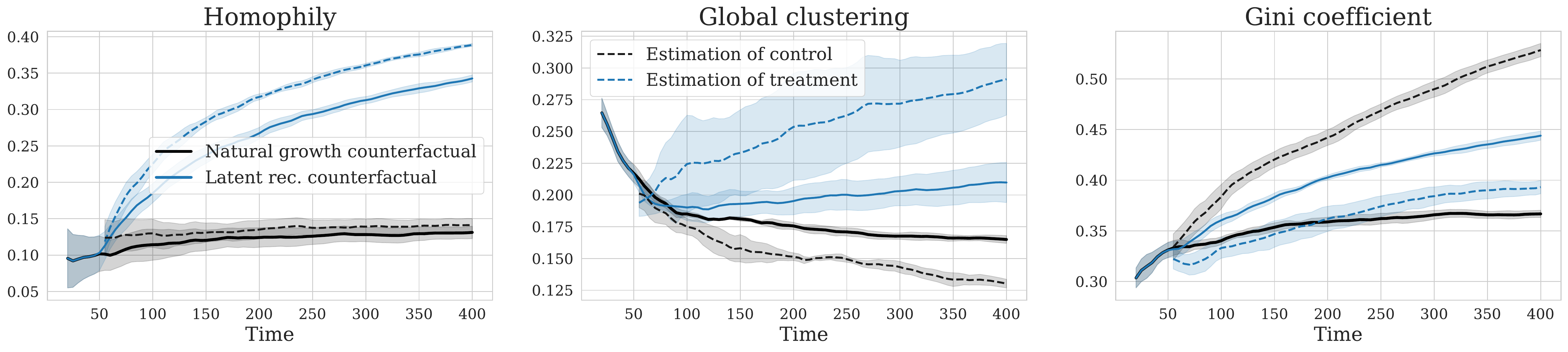}
    \caption{\latent{} recommender}
    \end{subfigure}

    \begin{subfigure}[b]{0.9\linewidth}
    \centering
    \includegraphics[width=\linewidth]{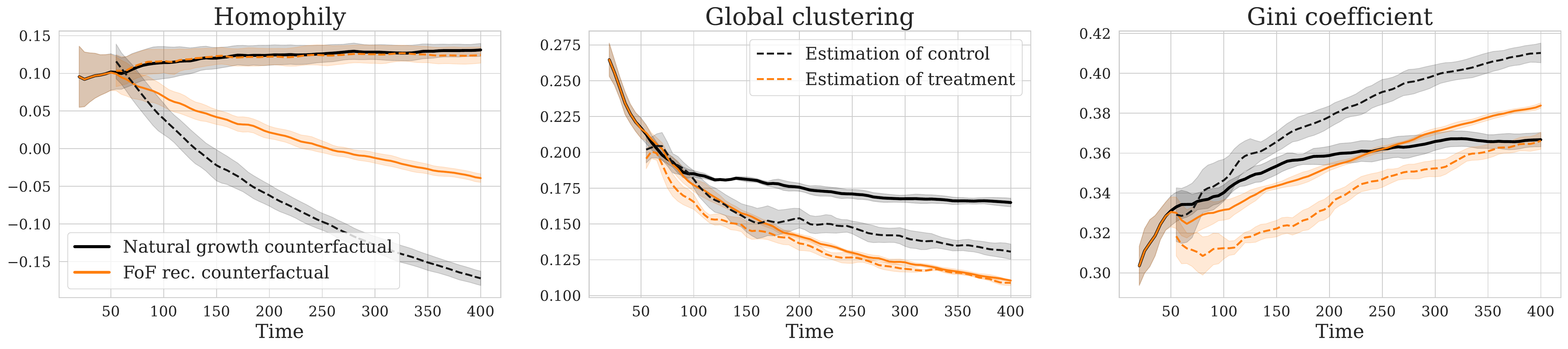}
    \caption{\fof{} recommender}
    \end{subfigure}
    \caption{A/B Evaluations: Solid lines correspond to ground truth counterfactual evaluation of homophily, clustering and Gini coefficient. Dashed lines correspond to trajectories estimated \emph{naively} from running an A/B test.}
    \label{fig:ab_naive}
\end{figure}

Fig. \ref{fig:ab_naive} shows the naive estimates derived from the A/B evaluation.

\end{document}